\documentstyle[12pt,psfig]{article}
\reversemarginpar
\begin{document}
\author{Thomas C. Bishop\footnote{Center for Bioenvironmental
Research at Tulane and Xavier Universities\newline  1430 Tulane
Ave, SL-3 New Orleans, LA, 70112\newline * V: 504-988-6203/F:
504-585-6145\newline * bishop@tulane.edu }, Oleksandr O.
Zhmudsky\footnote{Center for Bioenvironmental Research at Tulane
and Xavier Universities\newline Phone: (504)-585-6145,  E-mail:
ozhmuds@tulane.edu}}
\date{Keywords: chromatin, nucleosome, DNA, elastic rod, dynamics}
\title{Folding DNA into nucleosome and chromatin: dynamics.}
\maketitle
\begin{abstract}
  A theoretical framework for evaluating the approximate energy
and dynamic properties associated with the folding of DNA into
nucleosomes and chromatin is presented.  For this purpose
experimentally determined elastic constants of linear DNA and a
simple fold geometry are assumed to derive constants for the
higher order folding.  The model predicts the correct order of
magnitude for the experimentally determined Young's and shear
modulus of condensed chromatin.  Thus we have demonstrated that
the elastic properties of DNA are the primary determinant of the
elastic properties of each folded state.  The derived elastic
constants are then used to predict the speed of propagation of
small amplitude waves. It is shown that extension/compression,
twist, bend or shear waves can be excite in each folded state.
\end{abstract}
{\Large\bf Introduction}

The folding of DNA into higher order structures is readily
observed, but the path of DNA through these folded structures has
not been determined conclusively \cite{Holde}.  Only the structure
of individual nucleosome particles that have been crystallized is
available at atomic resolution \cite{Luger}.  The x-ray structures
of the nucleosome reveal that DNA is wrapped around a histone core
in a somewhat irregular, left-handed helical path, but these
structures do not provide information on how multiple nucleosomes
are arranged on a length of double stranded DNA in chromatin.  For
this purpose a variety of experimental techniques (see
\cite{Holde}) as well as computational modeling
\cite{Katritch,Beard,Ehrlich,Westcott} have been used. In our
model the hierarchy of folding: DNA, nucleosomes, condensed
chromatin, corresponds to the hierarchy of equilibrium
conformations that exist for an elastic rod
\cite{ShiHearst,ShiBorovik,Shi}.  In elastic rod theory these
folded states are unstable, but biologically we know that the
first folding (nucleosomes) is stabilized by the histone octamer
and the second folding (condensed chromatin) is stabilized by
linker histone \cite{Thoma}.

Both an extended and condensed form of chromatin exists.  The
extended conformation of chromatin is an irregular, 3-dimensional
zig-zag pattern of what appears to be intact nucleosomes unevenly
spaced along DNA.  The degree of compaction of this zig-zag
structure is known to be affected by the ionic environment
\cite{Widom}. The length of linker DNA between nucleosomes and the
entry/exit angle of the linker from each nucleosome are important
but nonetheless secondary determinants. The primary determinant of
the structure of extended chromatin is the histone octamer, since
under appropriate conditions the zig-zag can be condensed into a
more or less continuous, irregular $11\;nm$ fiber with essentially
zero length linker.  To obtain the next level of folding, what we
refer to as condensed chromatin with a diameter of approximately
$30\;nm$, requires the presence of linker histones \cite{Thoma}.
Thus the primary determinant of the folding of extended chromatin
into condensed chromatin is linker histone. The path of DNA in
condensed chromatin is determined by the arrangement of
nucleosomes within the fiber and in turn, it is expected that the
arrangement of nucleosomes is determined by whether or not the
linker DNA between nucleosomes bends or remains straight.

A recent review \cite{Holde} favors straight linkers and an
arrangement of nucleosomes that resembles the 2-start model over
the solenoid \cite{Finch} or coiled-linker models, (the latter two
assume a bent linker), but the evidence is still inconclusive.
Virtually all models assume the nucleosome maintains the x-ray
structure. This does not seem justified when it is known that the
two (H2A-H2B) dimers that form part of the octamer core can
readily dissociate from the nucleosome in solution and can be
reversibly dissociated by changing ion concentrations
\cite{Eickbush}. Furthermore the affects of linker histone on the
structure of the nucleosome, in particular the histone core, are
not known.  A "pear-shaped" structure is apparently induced in the
nucleosome upon the binding of linker histones, as observed with
electron microscopy \cite{Fritzsche,Bednar} and scanning force
microscopy \cite{Fritzsche}. Single fiber pulling experiments
demonstrate that there are discrete "jumps" as histones dissociate
from extended chromatin fibers suggesting a simple on/off
association of the histone octamer, but the data also indicate a
linear force-extension relationship for the gross properties of
the extended chromatin fiber, Figure 3b in \cite{Bennink}.

Our view is that nucleosomes in particular is the arrangement of
the core histones will likely be altered by both the addition of
linker histones and the constraints imposed by folding or external
forces.  The latter occur during experimental manipulations, as
well as, biological processes.  The observed irregularities in
extended and condensed chromatin result from a fluid-like motion
of histones within nucleosomes and of nucleosomes along chromatin
in a fairly smooth energy landscape with multiple local minima.
The local minima are due in part to the sequence dependent nature
of DNA, the particular state of modification of the DNA and the
histones, and differences in the local environment.  At the
molecular level, the local environment has an inhomogeneous
distribution of ions and/or DNA binding proteins.  The barrier
between these minima is relatively low thus well-defined regular
structures do not exist. We assume that to a first approximation,
i.e. we smooth out the energy landscape even further, the
structures of extended and condensed chromatin are governed by the
elastic properties of DNA and can be predicted based on the theory
of elastic rods \cite{ShiHearst,ShiBorovik,Shi,Bishop,Hearst}. As
noted by Bishop and Hearst \cite{Bishop}, the "pear-shaped" or
elliptic structures that are observed are a tell-tale sign that
DNA is behaving as an elastic rod even when it is folded into
nucleosomes and the energy landscape associated with a potential
function describing chromatin folding is sufficiently flat to
allow for the irregularities noted above. Furthermore the effect
of combining a histone core that tends towards cylindrical
symmetry with a wrapping of DNA that tends towards elliptic
cross-sections, as predicted by elastic rod theory, produces an
interaction that explains the deviation in the path of DNA as it
passes over the dyad axis of symmetry of the nucleosome.

In order to extend our potential function representation of the
folding of DNA into nucleosomes and chromatin, we seek first to
determine to what degree DNA alone contributes to the observed
properties of these structures.  For this purpose we analyze the
elastic properties of an idealized nucleosome that results from
folding DNA into a regular left handed helix.  If this helix is
extended beyond the $1.75$ turns that correspond to a single
nucleosome, it will form a linear array of nucleosomes aligned end
to end, as has been suggested for the structure of telomeric DNA
\cite{Fajkus}. We propose that such a model approximates the
properties of extended chromatin in the limit of a zero length
linker even if it does not represent the actual structure
\cite{Remark}.  In a like manner condensed chromatin is modelled
as a regular right handed helix constructed from a linear array of
nucleosomes, similar to the solenoid or coiled linker models and a
close approximation to the helix-on-a-linear helix predicted by
elastic rod theory.  The model predicts that the elastic
properties of DNA are the primary determinant of the elastic
properties of each folded state.  Our results can also be used to
relate experimentally determined elastic properties to the
velocity of propagation of mechanical disturbances in DNA and
chromatin, which has direct relevance for molecular processes.

In the next section we present the equations of motion of an
elastic rod and outline the derivation of our linear analysis
\cite{Zhmudsky}.  We then demonstrate how to use the known elastic
constants of DNA to evaluate the velocity of propagation of
various mechanical disturbances through DNA.  In subsequent
sections we use the elastic constants for DNA to obtain effective
elastic constants for nucleosomes, extended and condensed
chromatin.  For each fiber we calculate the speed of propagation
of mechanical disturbances.

   Our paper is based on a continuous medium model of DNA.
Such an approach is well known and widely used to describe solids,
liquids and gases (see, for example \cite{Landau1,Landau2}). Thus,
when we speak of infinitely small elements of volume, we shall
always mean those which are "physically" infinitely small, i.e.
very small compared with wavelength or radius of curvature under
consideration, but large compared with the distance between the
atoms in DNA.

\section{Dynamics of an Elastic Rod.}

We utilize a uniform isotropic elastic rod model for which
analytic solutions for the equilibrium configurations have been
determined \cite{ShiHearst} and parameterized to represent the
observed folding of DNA \cite{Hearst}.  Such an elastic rod model
is a suitable description for small deformations of any solid body
that is long, slender and possesses uniform material properties.
Here, small refers to the length scale of the deformation in
comparison to the length scale of the rod, but still large
compared to the distance between atoms.  Thus the results
presented in this section apply to macroscopic objects such as
cables and beams, as well as hair and cilia, actin filaments and
DNA. Strictly speaking all experiments cited above silently
suppose that DNA is a continuous media with no structure. One only
needs to utilize the appropriate elastic constants for each entity
to apply the results.  The equations of motion \cite{Simo} for
such an elastic rod can be solved numerically and visualized in
3-dimensional space \cite{McClain}.

A natural coordinate system for expressing the equations of motion
of an elastic rod is a so-called internal coordinate system that
relates the translation, denoted by the three-vector $\vec\Omega$,
and the rotation, denoted by the three-vector, of one
cross-section with respect the adjacent one. These six coordinates
have a one to one correspondence to the six helical parameters
that describe DNA. In biologic terminology the three components of
$\vec\Gamma$ correspond to the basepair parameters (shift, slide,
rise) and the three components of $\vec\Omega$ correspond to
(roll, tilt, twist). Formally, the components of $\vec\Gamma$ and
$\vec\Omega$ are orthogonal and are defined only in the limit of
infinitesimally thin "basepairs". For an actual strand of DNA,
$\vec\Gamma$ and $\vec\Omega$ will also be functions of time, $t$,
and location along the strand of DNA, $s$. Each basepair will also
possess translational and rotational velocity, denoted by
$\vec\gamma$ and $\vec\omega$. The equations of motion expressed
in such a coordinate system are as follows:
\begin{eqnarray}
\rho\left({\partial\vec\gamma\over\partial t}+\vec\omega\times
\vec\gamma \right)&=&\hat
C\cdot{\partial(\vec\Gamma-\vec\Gamma_0(s)) \over\partial s}+\vec
\Omega\times\left(\hat C\cdot\left(\vec\Gamma-\vec\Gamma_0\right)\right) \label{intro:1}\\
\hat I\cdot{\partial\vec\omega\over\partial t}+\vec\omega\times
\left(\hat I\cdot\vec\omega\right)&=&\hat
D\cdot{\partial(\vec\Omega-\vec\Omega_0(s)) \over\partial s}+\vec
\Omega\times\left(\hat
D\cdot\left(\vec\Omega-\vec\Omega_0\right)\right)\nonumber\\ &+&
\vec\Gamma\times\left(\hat
C\cdot\left(\vec\Gamma-\vec\Gamma_0\right)\right) \label{intro:2}\\
{\partial\vec\Gamma\over\partial t}+\vec\omega\times
\vec\Gamma&=&{\partial\vec\gamma\over\partial s}+\vec\Omega\times
\vec\gamma
\label{intro:3}\\
{\partial\vec\Omega\over\partial t}+\vec\omega\times
\vec\Omega&=&{\partial\vec\omega\over\partial s}
\label{intro:4}\end{eqnarray}

  The diagonal matrix $\hat I$ is the linear density of the inertia
tensor for a cross-section of the rod.  The matrix $\hat C$
contains the shear ($\mu$) and Young's modulus ($Y$) as the
diagonal elements and zeros everywhere else; $\hat D$ contains the
bend stiffness ($D_{1,2}$) and torsional rigidity ($D_3$) as
diagonal elements; and $\rho$ is the linear density of the rod.

The first equation in (\ref{intro:1}-\ref{intro:4}) represents the
balance of force and linear momentum and the second equation
represents the balance of torque and angular as written in the
reference frame of the rod. The use of Hooke's Law to describe
bending/twisting deformations with respect to some arbitrary
intrinsic bend/twist state denoted by $\vec\Omega_0$ is apparent.
A similar statement applies to shear/stretch deformations with
respect to some intrinsic shear/extension  $\vec\Gamma_0$. For
ideal B-from DNA $\vec\Omega_0=(0,0,36^0/bp)$ and
$\vec\Gamma_0=(0,0,3.4\;\r A/bp)$. The third and fourth equations
arise from geometrical considerations. More thorough descriptions
can be found in \cite{Simo,McClain}.

In subsequent sections we consider DNA as an elastic rod, extended
chromatin as an elastic rod and condensed chromatin as an elastic
rod.  In each case we only have to determine appropriate values
for each of the matrices in the above equations.  Once these
values have been determined the velocity of propagation of a
mechanical disturbance through each structure can be evaluated as
described below.

A linear analysis (small amplitude disturbances) of equations
(\ref{intro:1}-\ref{intro:4}) indicates that four different types
of waves can propagate through the elastic rod \cite{Zhmudsky}.
These are an extension/compression, bend, twist, or shear waves in
the rod. In the limit of very short wavelength (wavelength is much
smaller than scale parameters of the problem, e.g. much less than
radius of curvature of the rod) these four types propagate
independently of one another and independently of a shape of a
rod. In this case expressions for speed of waves propagation will
be obtained below.

The linear analysis assumes that the four three-vector functions
$\vec\Gamma(s,t)$, $\vec\Omega(s,t)$, $\vec\gamma(s,t)$ and
$\vec\omega(s,t)$ each have the functional form  $\vec
G_i(s,t)=\vec G_{i0}\cdot\exp(-i\omega t+iks)$ where $\vec G_{i0}$
is a constant. Upon substitution into the linearized equations of
motion one obtains the following relations:
\begin{eqnarray}
i\omega^2\rho\vec\Gamma-ik^2(\hat C\cdot\vec\Gamma)&=& 0
 \label{linear:4i}\\
i\omega^2 (\hat I\cdot\vec\Omega) - ik^2\left(\hat
D\cdot\vec\Omega\right) &=& 0 \label{linear:4j}
\end{eqnarray}
It is convenient to split the longitudinal and transverse
components with help of the definitions
$\vec\Gamma_\bot\equiv(\Gamma_1,\Gamma_2,0)$ and
$\vec\Omega_\bot\equiv(\Omega_1,\Omega_2,0)$. In this way we can
easily divide the four types of wave types which can propagate
along the rod. The equations below correspond to four waves:
shear, extension, bend and twist respectively:
\begin{eqnarray}
(\omega^2\rho - k^2 C)\vec\Gamma_\bot &=& 0
\label{linear:4k}\\
(\omega^2\rho - k^2 C_3)\Gamma_3 &=& 0
\label{linear:4l}\\
(\omega^2 I - k^2 D)\vec\Omega_\bot &=& 0 \label{linear:4m} \\
(2\omega^2 I - k^2 D_3)\Omega_3 &=& 0 \label{linear:4n}
\end{eqnarray}
We point out that in general case (arbitrary wavelength) bend and
shear waves are coupled but extension and twist remain
independent. Moreover it can be shown that for straight rod
extension and twist waves have "special" properties. They are
independent of each other and of bend and shear waves even for not
small wave amplitudes. Our assumption of a circular cross section
($D_1=D_2$, $C_1=C_2$ and $I_{xx}=I_{yy}$) results in a
equivalents speed of propagation for both components of bend and
shear.

Explicit expressions for the wave velocities are:
\begin{itemize}
  \item Shear waves ($\vec\Gamma_\bot$):
\begin{equation}\label{dynamics:2}
 V_{Shear}=\sqrt{\mu\over\rho}
\end{equation}
  \item Extension waves ($\Gamma_3$):
\begin{equation}\label{dynamics:3}
 V_{Extension}=\sqrt{Y\over\rho}
\end{equation}
  \item Bend waves ($\vec\Omega_\bot$):
\begin{equation}\label{dynamics:4}
 V_{Bend}=\sqrt{D_{1,2}\over I}
\end{equation}
  \item Twist waves ($\Omega_3$):
\begin{equation}\label{dynamics:5}
V_{Twist}=\sqrt{D_3\over 2I}
\end{equation}
\end{itemize}

We point out that expressions (\ref{dynamics:2}-\ref{dynamics:5})
are rightly for a rod of arbitrary shape because in the short wave
limit all terms that define the rod shape in equations
(\ref{intro:1}-\ref{intro:4}) were vanished.

It is well known that the measurement of wave velocities is a
usual method for determining elastic properties of solids. Just so
the DNA elastic properties were studied in \cite{Hakim}.

\section{Dynamics of linear DNA.}\label{section:1}

Having obtained expressions (\ref{dynamics:2}-\ref{dynamics:5}) we
evaluate the propagation velocities of each type of mechanical
disturbance for linear B-form DNA using the elastic properties
listed in Table \ref{Table:1}. The results are listed in Table
\ref{Table:3}. For our calculations we have used a linear density
of DNA of $\rho_{dna} = 660\;Da/basepair\cdot basepair/0.34\r A$,
and assumed a DNA radius of $1.0\;nm$ with a circular
cross-section for determining the moment of inertia tensor.
\begin{table}[h]
\begin{tabular}{|l||c|l|}\hline
 \rule{0pt}{14pt} Elastic Constant  & Symbol & Value \\ \hline
 \rule{0pt}{14pt} Young's modulus & $Y$ & $1.09\cdot 10^{-9} [KM/S^2]$ \\ \hline
 \rule{0pt}{14pt} Shear modulus & $C$ & $8.16\cdot 10^{-9} [KM/S^2]$ \\ \hline
 \rule{0pt}{14pt} Torsion rigidity & $\mu$ & $2.02\cdot 10^{-28} [KM^3/S^2]$ \\ \hline
 \rule{0pt}{14pt} Bend stiffness & $A$ & $2.7\cdot 10^{-28} [KM^3/S^2]$ \\ \hline
 \rule{0pt}{14pt} Linear density & $\rho$ & $3.22\cdot 10^{-15} [K/M]$ \\\hline
 \rule{0pt}{14pt} Moment of inertia & $I_{xx,yy}$ & $8.05\cdot 10^{-34} [KM]$ \\
\rule{0pt}{14pt}  & $I_{zz}$ & $1.61\cdot 10^{-33} [KM]$
\\\hline
\end{tabular}\caption{Elastic Constants for linear DNA
\cite{Bouchiat,Moroz,Hakim, Cui}. } \label{Table:1}\end{table}

Here we have used the quantities $Y^* = Y\cdot A$ where $Y$ is the
Young's modulus as expressed in the text, $A$ is the area of the
cross-section, and $Y^*$ is the stretch modulus. Similarly $\mu^*
= \mu\cdot A$.

\section{Effective Elastic Constants for Nucleosomes and Chromatin.}

In this section we demonstrate how to determine effective elastic
constants for nucleosomes and condensed chromatin.  For this
purpose we analyze the nucleosome as a regular helical spring made
of DNA and condensed chromatin as a regular helical spring made of
a linear array of nucleosomes.  In case of the "nucleosome spring"
we consider two limiting cases.  In the first case, we imagine
that the histones stably fold the DNA into a helix, yet the DNA is
free to undergo small deformations independent of the histones.
For this case the histones are completely ignored, and the
nucleosome is really a spring made from DNA.  In the second case
we image that the DNA is rigidly attached to the histones but that
the elastic properties of the nucleosome are still governed by
DNA. In this case the nucleosome is a spring made from DNA but it
is filled with a core material made of histones.  For the analysis
we need to know the elastic constants of DNA, the linear density
of DNA and the histones and the fold geometry.  Since each fold is
a simple helix, the pitch and radius of the helix define the
geometry.  The radius of each helix is defined as the centerline
path of the DNA in case of the nucleosome and the center line path
of the nucleosomes in case of condensed chromatin.

\subsection{DNA elastic properties.}

\subsubsection{Geometry of folding.}

A length of linear DNA we shall designate by an $l$, a length of a
linear array of nucleosomes by an $L$ and a length of the
condensed chromatin by an ${\cal L}$.  The radius of linear DNA is
$a = 1.0\;nm$, the radius of the nucleosome is $R=4.5\;nm$ and the
radius of condensed chromatin is ${\cal R}=9.5\;nm$.  The
nucleosome's pitch is $h=2.5\;nm$ and chromatin's pitch is $H=15\;
nm$. Note that there are $1.75$ turns of the helix for a single
nucleosome, but for a linear array of nucleosomes there will be
$n$ turns.  Thus we can evaluate $L=n\cdot h$ and similarly for
$N$ turns of the $30 nm$ fiber ${\cal L}=N\cdot H$. The length of
linear DNA contained within $n$ turns of a linear array of
nucleosomes is $l_{nuc}=\sqrt{(n2\pi R)^2+(nh)^2}$ and the length
of a linear array of nucleosomes contained within $N$ turns of
condensed chromatin is $L_{cc}=\sqrt{(N2\pi {\cal R})^2+(NH)^2}$ .
This yields a factor of $l_{nuc}/L\sim 10$ for the compaction of
DNA into nucleosomes and an additional factor of $L_{cc}/{\cal
L}\sim 4$ for the compaction of nucleosomes into condensed
chromatin.

\subsubsection{Linear mass density and moments of inertia.}

In the limit that the DNA functions independently of the histones,
the linear density of the nucleosome is easy to calculate.  The
linear density of our "nucleosome spring" is :
\begin{equation}\label{moment:1}
  \rho_{nuc}=\rho_{dna}\cdot{l_{nuc}\over L}\approx 3.66\cdot
  10^{-14} [K/M]
\end{equation}
In the limit that the histones are rigidly attached to the DNA the
linear density of the nucleosome is the sum of the linear density
expressed above and the linear density of histones
($108,500\;Da/nucleosome$).  In this limit $\rho_{nuc} = 1.1\cdot
10^{-13}[K/M]$.

The linear density of condensed chromatin is accordingly
$\rho_{cc} = \rho_{nuc}L_{cc}/{\cal L}$ with a value of $4.5\cdot
10^{-13} [K/M]$ for the limiting case of the histones rigidly
attached to the DNA. We point out that to vary the linker length
the value of $l_{nuc}$ should be changed accordingly.

For determining the moments of inertia we consider each spring as
a hollow elastic rod.  The walls are made of DNA in case of the
"nucleosome spring" and of nucleosomes in case of the "condensed
chromatin spring".  Again, the nucleosome spring has two limits.
In the limit that the DNA is rigidly attached to the histones, the
core of the spring is filled with histones and the moments of
inertia are calculated according to the expressions for rods. In
the limit that the DNA is independent of the histones, the
nucleosome spring is hollow, and the moments of inertia are
calculated as for a tube.  Analysis of the elastic constants in
the latter limit is formally equivalent to our analysis of a
condensed chromatin spring.

\subsection{Elastic constants for Nucleosomes and Extended Chromatin.}

\subsubsection{Extension/compression constant of the nucleosome.}
Following the analysis of springs presented in Elmore and Heald
\cite{Elmore}, when our "nucleosome spring" is stretched along the
axial direction the DNA in the nucleosome undergoes a twisting
deformation.  The twisting deformation of the DNA treated as a
solid cylinder is related to its shear modulus, thus the
nucleosome spring constant $k_{nuc} = F/\Delta X$ relates to the
shear modulus of the DNA, $\mu_{dna}$, as follows:
\begin{equation}\label{Young:0}
  k_{nuc}={\pi\mu_{dna} a^4\over2 R^2_{cent} l_{nuc}}
\end{equation}
Here $a$ is the radius of the DNA, $R_{cent}$ is the centerline
radius of the nucleosome spring and $l_{nuc}$ is the length of DNA
in the nucleosome.  We now treat the nucleosome spring as a
cylinder with outer and inner radii, $R_{out}$ and $R_{in}$ and
cross-sectional area, $A = \pi\{(R_{out})^2 - (R_{in})^2\}$.  The
Young's modulus for such a hollow cylinder is (Young's modulus is
related to the stretch modulus $Y^* = Y\cdot A$):
\begin{equation}\label{Young:3}
  Y_{nuc}={k_{nuc}L\over A}= {\mu_{dna}a^4 L\over 2R^2(R_{out}^2-R_{in}^2)l_{nuc}}
\end{equation}

In the first limiting case the DNA moves independently of the
histone core so we choose $R_{out}=R+a$ and $R_{in}=R-a$.  In the
second limiting case the DNA is rigidly attached to the histone
core so we choose $R_{out}=R+a$ and $R_{in}=0$.  The corresponding
values for the Young's modulus are given in Table 2. There also
one can find shear modulus, torsion rigidity and bend shtiffness
that will be evaluated below.

\subsubsection{Shear constant.}

Again, following the analysis of springs in \cite{Elmore}, we
determine the shear constant for the nucleosome by supposing that
a pair of equal and opposite axial torques $M$ is applied to the
ends of our nucleosome spring that has length $L$.  Such a
twisting distortion of the nucleosome causes the radius of the
nucleosome to increase or decrease thus bending the DNA.  So the
twist constant of our nucleosome spring is a function of the bend
stiffness of the DNA. The torsion constant for the nucleosome
spring is:
\begin{equation}\label{shear:1}
  k_{\varphi-nuc}={M\over\varphi}={2IY\over l_{nuc}}={\pi
Y_{dna}a^4\over 2l_{nuc}}
\end{equation}
where $\varphi$ is twist angle of the nucleosome, $I=(\pi/4)a^4$,
is the moment of the DNA cross-section, $Y_{dna}$ is the Young's
modulus of the DNA, and $l_{nuc}$ is the length of DNA in the
nucleosome.  We now consider the nucleosome as a cylinder with
outer and inner radii, $R_{out}$ and $R_{in}$ and write an
expression for the torsion constant as follows:
\begin{equation}\label{shear:2}
  {M\over\varphi}={\pi\over 2}\mu_{nuc}{R^4_{out}-R^4_{in}\over L}
\end{equation}
Here $\mu_{nuc}$ is shear modulus of the nucleosome, and $L$ is
its length. Equating these two expressions we obtain ($\mu^* =
\mu\cdot A$ as for the Young's modulus):
\begin{equation}\label{shear:3}
  \mu_{nuc}={Y_{dna}a^4L\over (R^4_{out}-R^4_{in})l_{nuc}}
\end{equation}
In the first limiting case the DNA moves independently of the
histone core so we choose $R_{out}=R+a$ and $R_{in}=R-a$. In the
second limiting case the DNA is rigidly attached to the histone
core so we choose $R_{out}=R+a$ and $R_{in}=0$.  The corresponding
values for the shear modulus are given in Table~2. In that table
units are the same as in the Table \ref{Table:1}.

\subsubsection{Torsional rigidity constant.}

Having obtained $\mu_{nuc}$ we can now evaluate the torsional
rigidity constant of a cylinder as a function of shear modulus:
\begin{equation}\label{Rigidity:1}
D_3^{nuc}=\frac 12\mu_{nuc}\pi (R^4_{out}-R^4_{in})
\end{equation}
In the limiting case that the DNA moves independently of the
histone core $R_{out}=R+a$ and $R_{in}=R-a$.  In the other limit
$R_{out}=R+a$ and $R_{in}=0$.  The values for the torsional
rigidity are listed in Table~2.

\subsubsection{Bend stiffness constant.}

To complete the analysis, the bending stiffness of a hollow rod
with circular cross-section is a function of its Young's modulus
as follows:
\begin{equation}\label{Rigidity:2}
D_{1,2}^{nuc}=\frac 14Y_{nuc}\pi (R^4_{out}-R^4_{in})
\end{equation}
In the the limiting case that the DNA moves independently of the
histone core $R_{out}=R+a$ and $R_{in}=R-a$.  In the other limit
$R_{out}=R+a$ and $R_{in}=0$.  The values for the bending
stiffness  are listed in Table~2.

\subsection{Elastic constants for Condensed Chromatin.}

 Since we treat condensed chromatin as a spring made from a linear
array of nucleosomes, which contain a solid core of histones, the
expressions for the elastic constants for condensed chromatin are
identical in form to the previous section.  The only differences
are that instead of using the pitch and radius of the nucleosome
and the elastic constants of DNA we use the pitch and radius for
condensed chromatin (i.e. $L$ is replaced everywhere by $L$,
$l_{nuc}$ by $L_{cc}$, and a by $R$) and elastic constants for the
nucleosome.  We only use the elastic constants for the nucleosome
obtained in the limiting case of DNA rigidly attached to the
histone core, and we analyze condensed chromatin with a hollow
core (i.e. $R_{out}={\cal R}+R+a$ and $R_{out}={\cal R}-R-a$ where
${\cal R}$, $R$ and $a$ are the centerline radii of condensed
chromatin and the nucleosome and the radius of DNA, respectively).
The resulting values for the elastic constants for condensed
chromatin are given in Table~2. Alternatively experimentally
determined values for extended chromatin can be used at this step
to determine elastic constants for condensed chromatin, or
experimentally determined elastic constants for condensed
chromatin can be used to determine effective constants for
extended chromatin.

\begin{table}[h]\label{Table:2}
\begin{tabular}{|l|c|l|c|}\hline
 \rule{0pt}{14pt} Elastic   & Extended  & Extended & Condensed\\
 \rule{0pt}{14pt} Constant  & Chromatin & Chromatin & Chromatin\\
 \rule{0pt}{14pt}           & (no core) & (histone core) & \\ \hline
 \rule{0pt}{14pt} Young's modulus & $1.8\cdot 10^{-12}$ & $1.8\cdot 10^{-12}$ & $9.2\cdot 10^{-14}$ \\ \hline
 \rule{0pt}{14pt} Shear modulus & $2.3\cdot 10^{-12}$ & $3.2\cdot 10^{-12}$ & $7.6\cdot 10^{-14}$\\ \hline
 \rule{0pt}{14pt} Torsion rigidity & $4.8\cdot 10^{-29}$ & $6.8\cdot 10^{-29}$ & $9.2\cdot 10^{-30}$ \\ \hline
 \rule{0pt}{14pt} Bend stiffness & $1.9\cdot 10^{-29}$ & $1.9\cdot 10^{-29}$ & $5.5\cdot 10^{-29}$\\ \hline
 \rule{0pt}{14pt} Linear density & $3.7\cdot 10^{-14}$ & $1.1\cdot 10^{-13}$ & $4.5\cdot 10^{-13}$\\\hline
 \rule{0pt}{14pt} Moment of inertia & $3.9\cdot 10^{-31}$ &  $6.1\cdot 10^{-31}$ &  $9.0\cdot 10^{-30}$\\
  \rule{0pt}{14pt} & $7.8\cdot 10^{-31}$ &  $1.2\cdot 10^{-30}$ &  $1.8\cdot 10^{-29}$\\\hline
\end{tabular}\caption{Elastic constants for extended and condensed
chromatin.}
\end{table}

In this table we use the same units as in Table \ref{Table:1}.

\section{Wave Propagation in Extended and Condensed Chromatin.}
In this section we use the derived elastic constants listed in
Table~2 to evaluate the dynamics of extended and condensed
suitable chromatin.  The elastic constants for the nucleosome are
for measuring the force and torque associated with distortions of
individual nucleosomes. Propagating of mechanical disturbance
through the single nucleosome will be problematic because of
length scale requirement. However we can consider the propagation
of waves through a linear array of nucleosomes as an approximation
to extended chromatin.

These results are valid for any shape in which the curvature of
the rod is much greater than the wavelength of the disturbance.
 The velocities of propagation of mechanical disturbances for extended
and condensed chromatin are listed in Table \ref{Table:3}.
\begin{table}[h]
\vspace{3mm}
\begin{tabular}{|l|c|c|c|c|}\hline
 \rule{0pt}{14pt} Velocity & Linear & Extended   & Extended & Condensed\\
 \rule{0pt}{14pt}          &  DNA   & Chromatin  & Chromatin & Chromatin\\
 \rule{0pt}{14pt}          &        &(no histone)& (histone) & (histone) \\ \hline
 \rule{0pt}{14pt} Shear $(\r A/ns)$ & $5000$ & $79$ & $46$ & $4.1$\\ \hline
 \rule{0pt}{14pt} Bend $(\r A/ns)$   & $5100$ & $110$ & $89$ & $5.1$\\ \hline
 \rule{0pt}{14pt} Twist $(\r A/ns)$  & $3600$ & $79$ & $63$ & $3.6$\\ \hline
 \rule{0pt}{14pt} Extension $(\r A/ns)$& $5800$ & $70$ & $41$ & $4.5$\\ \hline
\end{tabular}\caption{Shear, bend, twist and extension wave velocities.}
\label{Table:3}\end{table}

\section{Conclusions.}

We briefly compare the predictions of our model to published
experimental results.  First the speed of sound in B-form DNA as
measured by Brillouin scattering has been reported as $1.9 km/s$
\cite{Hakim}, which differs from the value listed in Table
\ref{Table:3} by a factor of 3.  Thus our calculations indicate
that experimental results obtained from fundamentally different
approaches agree quite well, and that relating the speed of sound
in DNA to its elastic properties (i.e. equations
(\ref{dynamics:2}-\ref{dynamics:5})) is valid for DNA as it is for
other macroscopic materials.

For the evaluation of the elastic properties of extended
chromatin, we note that there is a linear relationship between the
average stretching force and the extension of $\lambda$-DNA in
Figure 3b of \cite{Bennink}. The corresponding spring constant is
approximately $10pN/10\mu m$. The sample contains core histones
but no linker histones thus according to our model this system is
extended chromatin.  We note that the degree of compaction, $16um$
DNA$/2um$ chromatin, is in close agreement with our value of 10
for the extended chromatin model. Converting the force constant to
a Young's modulus using Equation \ref{Young:3} gives a value of
$Y^* = YA = 2\cdot 10^{-12} [KM/S^2]$ which is the same as listed
in Table~2 for extended chromatin.  Thus the gross properties of
the stretching are determined by the elastic properties of the
DNA, but the specific details remain dependent on the histone-DNA
interactions as illustrated by the results in Figure 3c of
\cite{Bennink}.

Our model of condensed chromatin yields a Young's modulus and bend
stiffness that closely predicts the persistence length and stretch
modulus measured for chicken erythrocyte chromatin \cite{Cui}.
This sample contained linker histone so the folding should
correspond to our model of condensed chromatin, however the
reported degree of compaction ($\sim 10$) corresponded to our
model of extended chromatin. The reported persistence length, $P =
30\;nm$, can be converted to a bending stiffness, $D_{1,2} = Pk_bT
=1.2\cdot 10^{-28}[KM^3/S^2]$ which differs by a factor of 20 from
the value listed in Table~2.  The reported stretch modulus of
approximately $Y^* = 5\cdot 10^{-12} N$ for condensed chromatin is
a factor of 50 different than listed in Table~2. These values are
actually much closer to our predicted values for extended
chromatin differing by factors of 2 and 3, respectively. We offer
two possibilities for explaining the discrepancy.  One is that the
degree of compaction reported in \cite{Cui} indicates that the
fiber should more closely resemble our model of extended chromatin
than condensed chromatin. The other possibility is that the
elastic properties of condensed chromatin are not determined
primarily by the elastic properties of DNA as we have assumed in
this manuscript, but that intermolecular interactions arising
during the folding must also be considered. In this regard, we
believe that our results which are based on a purely mechanical
model provide good agreement with experiment.

Our main point has been to demonstrate how the folding of DNA
leads to a hierarchy of time and energy as well as length scales.
For this purpose we have evaluated elastic constants that
correspond to ideal geometries of nucleosomes and extended and
condensed chromatin and then evaluated the speed of propagation of
mechanical disturbances through each of these structures.  The
hierarchy of lengths resulting from the folding of linear DNA into
nucleosomes provides a factor of 10 reduction in length and the
folding of nucleosomes into condensed chromatin provides another
factor of 4 in compaction of linear DNA.  Since the potential
energy of deforming an elastic body is directly proportional to
the elastic constants simple ratios highlight the hierarchy of
energy associated with the folding.  For example the Young's
modulus for DNA divided by the Young's modulus for extended
chromatin is approximately $10^5$ while the corresponding ratios
for torsional rigidity and bend stiffness are approximately $10$.
The velocities presented in Table \ref{Table:3} are an indication
of the hierarchy of time scales.  There is approximately a $100$
fold reduction in the velocity of a mechanical disturbance
propagating through DNA as compared to extended chromatin and less
than a factor of $10$ reduction between extended and condensed
chromatin. Assuming that similar ratios occur for yet even higher
order folding, the atomic scale of DNA can be folded into the
macroscopic scale of the cell in as few as $6$ or $7$ folds.  This
is the same order of folding that exists for chromosomes during
mitosis.

We emphasis that our model is an approximate model, it does not
necessarily correspond to the atomic structure of any of the
modeled structures or the sequence dependent nature of DNA.
Experimentally measured elastic constants that correspond to a
particular sample source and specific experimental conditions
should be utilized in our velocity expressions to determine the
speed of propagation of mechanical disturbances through extended
and condensed chromatin.  These speeds of propagation are of
fundamental importance because they are literally the speed of
sound in DNA.  For comparison the speed of sound in water
approximately $\sim 1.5 km/s$ ($15\r A/ps$) and in steel is $\sim
5 km/s$ ($50\r A/ps$). Mechanical disturbances associated with
biological processes propagate at the speed of sound whether
through the nucleus (water), DNA, or chromatin.  Knowing these
speeds of propagation enables us to identity through which medium
a mechanical disturbance may be propagating.

The fact that twist disturbances propagate along DNA is readily
observed during transcription and is implicit in DNA topological
studies of twist relaxation.  Disturbances exciting a bend, shear
or extension are not as readily identifiable, but certainly the
interaction of a DNA binding protein deep in the major groove of
DNA will affect such disturbances.

\section{Acknowledgement.}

This work was conducted in the Theoretical Molecular Biology
Laboratory at the Tulane-Xavier Center for Bioenvironmental
Research which was established under NSF Cooperative Agreement
Number OSR-9550481/LEQSF (1996-1998)-SI-JFAP-04 and supported by
the Office of Naval Research (NO-0014-99-1-0763).

\newpage
\tableofcontents

\begin{thebibliography}{30}
\bibitem{Remark} If we consider extended chromatin as consisting of nucleosome
core particles connected via linker DNA the effective model in our
approach is nucleosome springs connected via straight rods.  The
stretching force constant for the straight rod will be greater
than for the nucleosome spring, thus when the fiber is pulled only
the springs will extend.  The straight rod segments only serve to
lengthen the fiber.  A similar statement holds for our model of
condensed chromatin in which the stiffer spring consists of
segments of extended chromatin.
\end{thebibliography}
\end{document}